\documentclass{article}

    \PassOptionsToPackage{numbers, compress}{natbib}

    \usepackage[final]{neurips_2021}

\usepackage[utf8]{inputenc} 
\usepackage[T1]{fontenc}    
\usepackage{hyperref}       
\usepackage{url}            
\usepackage{booktabs}       
\usepackage{amsfonts}       
\usepackage{nicefrac}       
\usepackage{microtype}      
\usepackage{xcolor}         
\usepackage[toc,page]{appendix}
\usepackage{graphicx}
\usepackage{soul}
\usepackage{subfig}

\title{Fourier Neural Operator for Plasma Modelling}

\author{
  Vignesh Gopakumar \\
  UK Atomic Energy Authority\\
  Oxford, OX14 3EB \\
  \texttt{vignesh.gopakumar@ukaea.uk} \\
  \And
  Stanislas Pamela \\
  UK Atomic Energy Authority \\
  Oxford, OX14 3EB \\
  \texttt{stanislas.pamela@ukaea.uk} \\
  \And
  Lorenzo Zanisi \\
  UK Atomic Energy Authority \\
  Oxford, OX14 3EB \\
  \texttt{lorenzo.zanisi@ukaea.uk} \\
  \And
   MAST Team\\
  UK Atomic Energy Authority \\
  Oxford, OX14 3EB \\
  \AND
  Zongyi Li \\
  Caltech \\
  California, 91125 \\
  \texttt{zongyili@caltech.edu} \\
  \And
  Anima Anandkumar \\
  Caltech \\
  California, 91125 \\
  \texttt{anima@caltech.edu} \\
}

\begin{document}

\maketitle

\begin{abstract}
    Predicting plasma evolution within a Tokamak is crucial to building a sustainable fusion reactor. Whether in the simulation space or within the experimental domain, the capability to forecast the spatio-temporal evolution of plasma field variables rapidly and accurately could improve active control methods on current tokamak devices and future fusion reactors. In this work, we demonstrate the utility of using Fourier Neural Operator (FNO) to model the plasma evolution in simulations and experiments. Our work shows that the FNO is capable of predicting magnetohydrodynamic models governing the plasma dynamics, 6 orders of magnitude faster than the traditional numerical solver, while maintaining considerable accuracy (NMSE $\sim  10^{-5})$. Our work also benchmarks the performance of the FNO against other standard surrogate models such as Conv-LSTM and U-Net and demonstrate that the FNO takes significantly less time to train, requires less parameters and outperforms other models. We extend the FNO approach to model the plasma evolution observed by the cameras positioned within the MAST spherical tokamak. We illustrate its capability in forecasting the formation of filaments within the plasma as well as the heat deposits. The FNO deployed to model the camera is capable of forecasting the full length of the plasma shot within half the time of the shot duration.
\end{abstract}

\section{Introduction}

Predicting the evolution of plasma dynamics is crucial to building sustainable fusion devices as it allows us to anticipate the plasma behaviour and take real-time decisions on the trajectory of engineering input control parameters. In fusion research, prediction of future plasma states is mostly undertaken using numerical modelling codes such as BOUT++ \cite{Riva2019}, JOREK \cite{Hoelzl2021jorek}, JINTRAC \cite{jintrac} and SOLPS \cite{WIESEN2015480}. However, these codes are computationally intensive, taking weeks if not months for effective simulations to be run on high-performance computers \cite{Riva2019}. They are also often complex in terms of software infrastructure and advanced solver library dependencies, making them difficult to deploy for rapid iterative modelling.

Machine learning has offered an alternative data-driven route for obtaining quick, inexpensive approximations to numerical simulations \cite{simintelligence}. These machine learning models are often physics-informed allowing us to create networks that are interpretable and can generalise \cite{Karniadakis2021}. Over the past few years, the research scope has expanded to explore machine learning based models aimed at plasma physics\cite{van_de_Plassche_2020} \cite{Gopakumar_2020}. We build upon this existing work and extend it with applications of neural operator learning to model plasma within fusion devices. 

Recently, a new class of data-driven models known as neural operators aim to directly learn the solution operator of the PDE in a mesh-invariant manner \cite{kovachki2021neural,li2022physics}. Unlike standard deep learning models, such as convolutional neural networks from computer vision, neural operators are invariant to discretization and hence, better suited for solving PDEs. They generalize the previous finite-dimensional neural networks to learn operator mapping between infinite-dimensional function spaces. Examples include Fourier Neural Operator (FNO)~\citep{li2021fourier}, graph neural operator \cite{li2020multipole}, 
and DeepONet \citep{lu2019deeponet}.  We consider FNO in this paper due to its superior  cost-accuracy tradeoff~\citep{de2022cost}. 

Our work demonstrates the performance of Fourier Neural Operator models \cite{li2021fourier} on solving magnetohydrodynamic (MHD) equations, and benchmark it against more traditional surrogate model architectures such as a Convolutional-LSTM and U-Net. In addition to numerical modelling, we demonstrate that FNOs can be trained to predict the evolution of experimental plasmas as observed by High Speed cameras on the MAST spherical tokamak \cite{MAST_CCFE}. The ability to predict future plasma states based on camera imaging, as well as other plasma diagnostics, could allow us to design efficient control feedback loops for future fusion reactors. To the best of our knowledge, this the first time that Neural Operator learning has been applied to plasma scenarios. 

\section{Problem Setting}
\subsection{MHD Simulations of Plasma Blobs}
\label{section: mhd}
The first dataset used to explore the FNO method is a simplified Magneto-hydrodynamics (MHD) model in toroidal geometry. In practice, advanced simulations of plasma instabilities are run with the JOREK code, using complex MHD models, and realistic geometries \cite{Hoelzl2021jorek}, however for this paper a simplified setup was used as a proof-of-principle. The dataset is obtained using the JOREK code \cite{Hoelzl2021jorek}, with a physics model similar to the Reduced-MHD model described in \cite{Hoelzl2021jorek}, but with electrostatic and isothermal constraints, such that only the density $\rho$, the electric potential $\Phi$, and the toroidal vorticity $w$ fields are evolved. In toroidal geometry, if a toroidally axisymmetric density blob is initialised on top of a low background density, in the absence of any plasma current to hold the density blob in place, the pressure gradient term in the momentum equation acts as a buoyancy effect and causes the blob to move outwards. 

This is used to create a dataset of density blobs moving radially inside a toroidally axisymmetric slab grid, bounded by Dirichlet conditions. The blobs evolve until they reach the wall and eventually disappear through both convective mixing and diffusion. The density diffusion and viscosity are fixed, while the blob’s initial conditions are varied, both in position, width, and amplitude. A set of 120 simulations are run (100 for training, 20 for testing), all with the same time-duration achieved with 1000 simulation timesteps, which allows for the blobs to travel all the way to the boundary and diffuse almost entirely. The number of simulation timesteps are down-sampled to 100 slices for the training set. Likewise, the 2D poloidal grid resolution, which is 200$\times$200 bi-cubic finite-elements, is down-sampled to a regular grid of 100$\times$100. The simulations were run on the EOSC-hub cloud, using 16 cores per run, and each simulation took approximately 160 core hours.

MHD modelling is of importance as it helps us understand the evolution of filaments that form across the flux surfaces of the tokamak as shown in the right of figure \ref{fig: elms} \cite{Smith_2020}. Accurately capturing the dynamics and evolution of the filaments help us characterise Edge-Localised Modes (ELMs), a nonlinear instability. ELMs can contribute damage to the reactor wall, potentially initiate other instabilities and it is crucial to be able to gain control over them. Through the course of this paper, the simulation modelling we undertake is that of a relatively simple physics structures, designed to serve as a proof-of-concept, outlining the utility of the FNO in modelling MHD scenarios. These simulations modelled in figure \ref{fig: fno_mhd}, show the diffusion of a blob which are a simplified scenario as that of the full ELM simulations as shown on the right of figure \ref{fig: elms}.

\subsection{Fast-Camera Images of the MAST Tokamak}
The Mega Ampere Spherical Tokamak (MAST) was a nuclear fusion device situated at the Culham Centre for Fusion Energy in the UK, where initial set of experiments were demonstrated on a Tokamak design with a narrow aspect ratio, demonstrating better efficiency over the magnetic field control \cite{Chapman_2015}. MAST has now been succeeded by MAST-Upgrade (MAST-U), which is currently operational and experimented on for better control strategies and design of the exhaust for potential future reactors \cite{MAST_CCFE}.

\begin{figure}
    \centering
    \includegraphics[width=3.7cm]{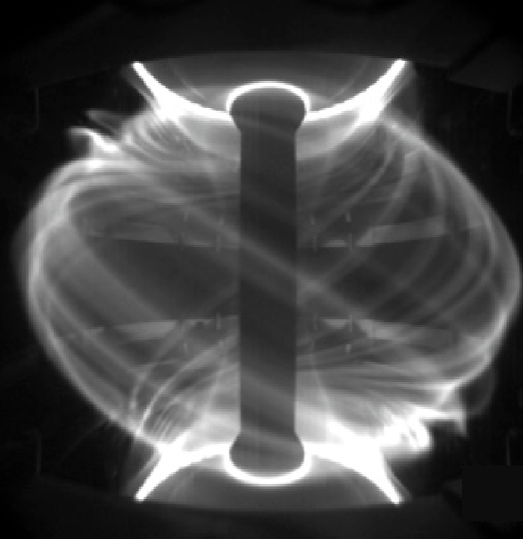}
    \label{fig: mast_fastcam}
    \includegraphics[width=8.0cm]{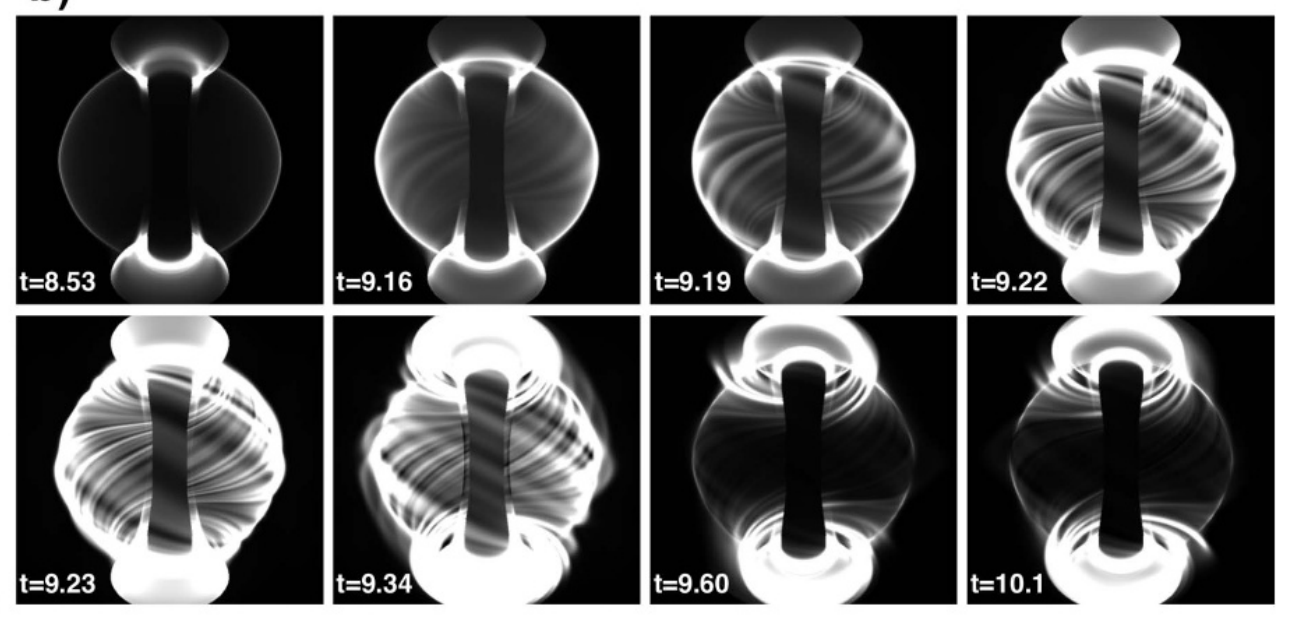}
    \label{fig: elms_siobhan}%
    \caption{Left: Fast camera image of a MAST plasma during an Edge-Localised-Mode, where plasma filaments are ejected outwards due to MHD instabilities at the plasma edge (figure reproduced from \cite{McArdle_MAST_2010}). Right: The evolution of the filamentary structures during the multi-mode ELM simulation imaged with a synthetic fast camera diagnostic (time given in ms). Full-scale MHD simulations as shown in this figure demonstrates how MHD behaviours are captured by the visible camera.  \cite{Smith_2020} }
    \label{fig: elms}
\end{figure}

MAST was equipped with fast camera imagining diagnostics at several points across the Tokamak for capturing images within the visible spectrum, displaying the plasma evolution. These Photron cameras \cite{photron} are essential diagnostics that contributed to fundamental understandings of key plasma phenomena in tokamaks in recent years \cite{Kirk2006}. While most of the camera data has been used for qualitative analysis in the past, it can be exploited using more advanced methods to provide statistical insight into plasma turbulence \cite{Walkden2022}, as well as global instabilities (disruptions) \cite{Ham2022}. Within the scope of this work, we focus on two camera view, one that looks at the central solenoid and the plasma around that to the walls, providing a holistic view of the plasma, and the so-called divertor-view, looking at the bottom of the plasma, where energy and impurities travel from the plasma to the wall (the plasma exhaust region). Figure-\ref{fig: elms} shows a view of the main plasma as seen by the camera(left), together with a synthetic rendering (right) of the camera, used for the calibration simulating the MHD physics. The cameras have a temporal resolution of 1.2ms, with spatial resolution varying according to the desired calibration set up for each experiment. They are tuned to visible wavelength and therefore capture mostly the Balmer D$_{\alpha}$ light emitted mostly from the plasma edge. As seen on Figure-\ref{fig: elms}, the fast-visible cameras are able to capture complex dynamics of filamentary eruptions at the plasma edge, the so-called Edge-Localised Modes (ELMs) \cite{Kirk2006}.


\section{Methods}
\subsection{Fourier Neural Operator}

The FNO approach shown in \cite{li2021fourier} learns a mapping  between two infinite dimensional spaces from a finite collection of observed input-output pairs. This is achieved by  way of a Fourier layer, as displayed in \cite{li2021fourier}. FNO is constructed by stacking multiple Fourier layers on top of each other. Each Fourier layer is composed of two learnable weight matrices: $R$ and $W$. $R$ learns the mapping behaviour within the Fourier space, while $W$ learns the mapping required within the input Euclidean space. The weight matrix $R$ within the Fourier layer enables convolution within the Fourier space, allowing the network to be parameterised directly in it. The output of a Fourier Layer can be expressed as: 

\begin{equation}
    y = \sigma\bigg(\mathcal{F}^{-1}\big(R\mathcal{F}(x)\big) + Wx\bigg)
    \label{fno_eqn}
\end{equation}

where, 
$x$ is the input, $y$ the output and $\sigma$ is the nonlinear activation function. $\mathcal{F}$ and $\mathcal{F}^{-1}$ represents the Fourier and inverse Fourier transform respectively. 

In this work, we will use the \textbf{FNO-2d} configuration, introduced in \cite{li2021fourier}, where a 2-d Fourier Neural Operator that only convolves in space is deployed in an auto-regressive manner. The FNO takes in as its input the field values across a 2D grid for an initial set of time instances (T\_in) along with the grid discretisation in both dimensions. It outputs a set of later time instances (step) of the field value across the same grid. The outptut of the FNO, coupled with the later time steps from the input are then fed back in an auto-regressive manner to estimate further field evolution in time, until the desired time instance (T\_out) is reached. For further information regarding the theory behind operator learning as ascribed by the FNO and its implementation in various case settings, we invite the author to refer to the original work in \cite{li2021fourier}. We have chosen the 2D configuration of the FNO as this gives us flexibility in deciding the length of the temporal evolution. 

The training conditions of the FNO over MHD data is given in \ref{section: fno_over_mhd}. \\

\subsection{U-Net and Convolutional LSTM}
\label{section: convlstm_unet}
Initially demonstrated in \cite{UNet},  further demonstrated in \cite{JIANG2021103878} and further evolved in \cite{ufno}, U-Net architectures have been instrumental in the design and structure of surrogate models for a vast variety of scenarios involving spatio-temoporal evolution. U-Net architecture is obtained by combining convolutional auto-encoders \cite{autoencoders} with skip connections \cite{resnets} at the decoder. The layers at each level of the encoder are skip connected and concatenated with corresponding layers at the decoder. The architecture we deployed for modelling reduced MHD was inspired by the work demonstrated in the PDE benchmark dataset: PDEBench \cite{pdebench}. 

Initially demonstrated in \cite{ConvLSTM}, the Convolutional LSTM (ConvLSTM) replaces the fully-connected layers in a LSTM \cite{lstm} with convolutional layers. This allows for the representation of additional spatially dependent layers within the recurrent structure of the model. We devise a ConvLSTM consisting of 5 ConvLSTM cells and train under the same conditions as for the FNO mentioned in section \ref{section: fno_over_mhd}. For a detailed layout of the architecture deployed for the ConvLSTM, refer to Appendix \ref{appendix_convlstm}.

\section{Results}
\subsection{FNO to Predict MHD Simulations}
\label{section: fno_over_mhd}
We construct and train independent 2D Fourier Neural Operator to model the spatio-temporal evolution of density ($\rho$), potential ($\phi$) and vorticity ($w$) associated with the plasma blob experiment as described in section \ref{section: mhd}. The architecture of the FNO was borrowed from \cite{li2021fourier} and is further laid out in the appendix \ref{appendix_fno2d}. Each model was trained on a single A100 GPU at the Cambridge Centre for Data Driven Discovery (CSD3). 

Each FNO is trained to learn the map between 20 time instances (T\_in=20) of a given field to the next 5 time instances (step=5). The output 5 time instances are concatenated with its previous 15 time instances to form the next 20 time instances of the input. This auto-regressive temporal evolution of the field is carried on until the $70^{th}$ time instance (T\_out=70). Each Fourier layer is characterised by a width of 32 and learning up to 16 modes in the Fourier space. Prior to training, the training dataset, built as explained in in section \ref{section: mhd} is normalised using a linear range scaling to lie between -1 and 1 for each of the field variables. All models are trained using the Adam optimizer \cite{adam} for 500 epochs with a batch size of 5. The learning rate is initially set to 0.001 and scheduled to decrease by half after every 100 epochs. Relative LP loss function was utilised for training the model. 

\begin{figure}
    \centering
    \subfloat[Density]{{\includegraphics[width=13.0cm]{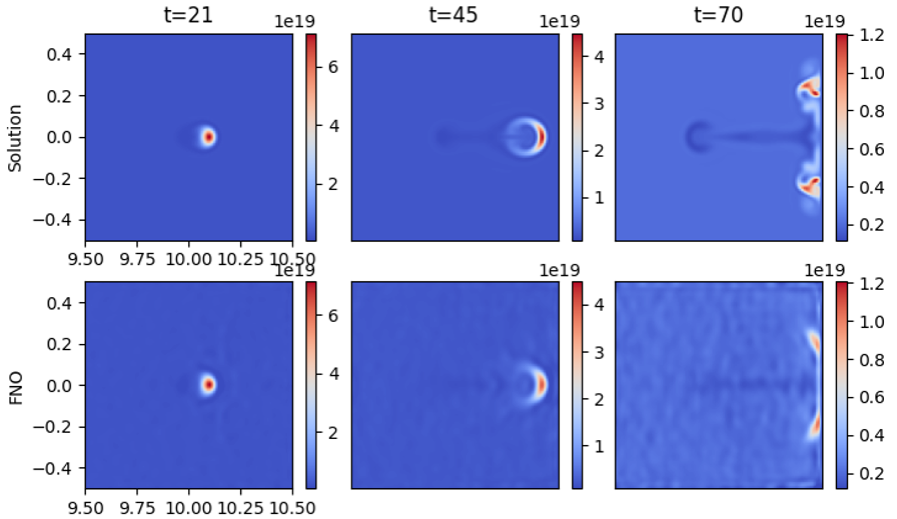}}
    \label{fig: fno_rho}}%
    \qquad
    \subfloat[Vorticity]{{\includegraphics[width=13.0cm]{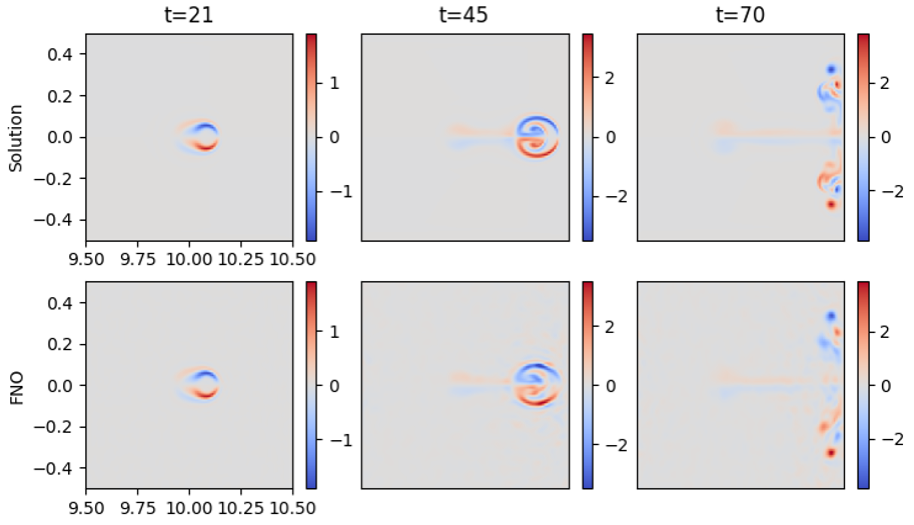}}
    \label{fig: fno_w}}%
    \caption{Temporal evolution of (a) the density and (b) the vorticity variables describing the plasma evolution as obtained using the JOREK code (top of each image) and that of the individually trained FNOs (bottom of each figure). The spatial domain is given in toroidal geometry characterised by R in the x-axis and Z in the y-axis.}
    \label{fig: fno_mhd}%
\end{figure}

As shown in figure \ref{fig: fno_mhd}, individual FNOs crafted to model the evolution of each field variable reproduce well the dynamics of the reduced MHD simulations. Figure \ref{fig: fno_rho} represents the time evolution of the density of the solution from JOREK and the FNO at the $21^{st}$, $45^{th}$ and $70^th$ time instance having been fed in the first 20 time instances. Figure \ref{fig: fno_w} shows the same for the evolution of the vorticity of the reduced MHD model (note we do not show the electric potential $\Phi$ in Figure-\ref{fig: fno_mhd}, since the vorticity $w$ is effectively the laplacian of $\Phi$). Our model achieved a Normalised MSE of 4.2$\times$10$^{-5}$ for mapping the evolution of electric potential, 5$\times$10$^{-5}$ for density and 4.6$\times$10$^{-5}$ for that of the vorticity. The FNO is able to evaluate the full solution in 0.025 seconds. Plots showing the contrast of the FNO output with that of JOREK by way of discrepancy plots can be found in Appendix \ref{appendix_errorplots}. \\

\textbf{U-Net and Conv-LSTM :} Both the U-Net and the ConvLSTM with achitectures mentioned in appendices \ref{appendix_unet} and \ref{appendix_convlstm} were trained under the same conditions as the FNO (as described above), and with the same auto-regressive strategy. Each model was trained on a single A100 GPU at the Cambridge Centre for Data Driven Discovery (CSD3). As demonstrated in table \ref{table: performance}, FNO takes significantly less time to train, requires less parameters and outperforms the ConvLSTM and the U-Net in modelling the diffusion of a blob of plasma described by reduced MHD. For conciseness we have only showed the results from the models mapping the evolution of potential in \ref{table: performance}, models deployed for learning other field variable have similar performance estimates, model composition and time consumption. For seeing the performance across each field variable and their comparative plots refer Appendix \ref{appendix_cross_model_performance}. 

\begin{table}
  \caption{Comparative Performance across various Models for Potential. Numerical Simulation ran on 10 cpu cores, training and inference of ML models on Nvidia A100 cards.}
  \label{table: performance}
  \centering
  \begin{tabular}{lllll}
    \toprule
    Model    &  Train Time (mins) & \# Parameters & NMSE & Inference Time(s)\\
    \midrule
    JOREK &   - & - & - & 36000 \\
    FNO &    13 & 2106981 & \textbf{4.245e-05} & 0.025    \\
    ConvLSTM  & 481 & 4874648 & 1.12e-03  &  0.14   \\
    U-Net     & 26 & 7931909 & 4.48e-04  & 0.033 \\
    \bottomrule
  \end{tabular}
\end{table}

\subsection{FNO to Predict Experimental Camera Data}
In order to create a data-driven digital twin capable of modelling the evolution of plasma as witnessed by the fast-cameras on MAST we invoked the 2D FNO architecture deployed in a auto-regressive framework as we had done in modelling the plasma described by reduced MHD. The FNO is well-suited for this task as the fast-camera images essentially capture the larger MHD behaviours exhibited by the experimental plasma. The utilisation of a Fourier based model is well justified as it allows to extract the modal behaviours within the plasma, consistent with it's periodic properties. The camera calibrated for specific experimental runs capture the information of the plasma onto a uniform pixel grid, allowing us to perform Fourier transformations over them. By using \textit{calcam}, a calibration tool developed by the European Atomic Energy Community \cite{scott_silburn_2022_6891504} combined with 3D CAD models of MAST, we are able to determine the domain range covered by each camera calibration. The upper and lower limits of the domain range of these images are then fixed onto uniform grids in both the R and Z axis, allowing us to create uniform grid discretisations underlying each camera image.

\begin{figure}
    \centering
    \includegraphics[width=\columnwidth]{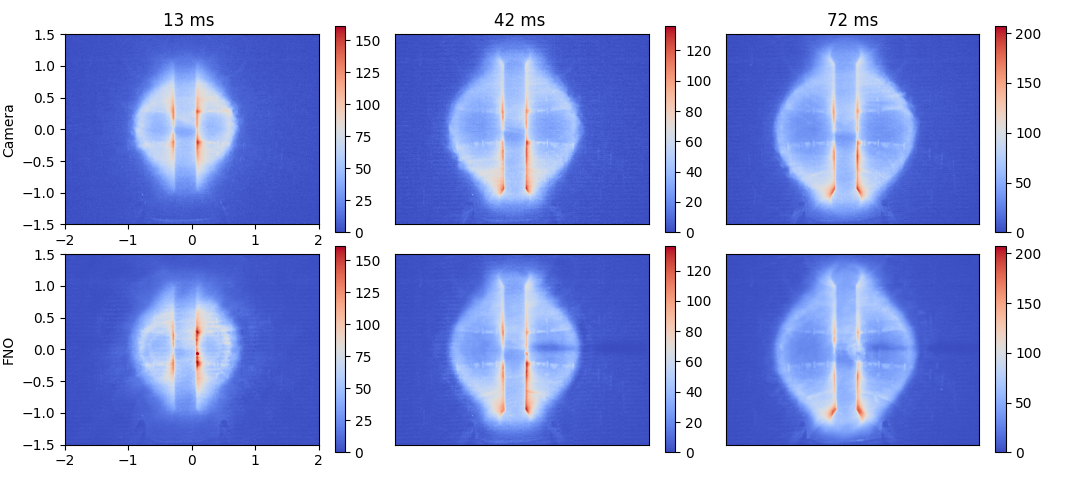}
    \caption{Temporal evolution of the plasma around the central solenoid within MAST as perceived by the main camera. The top layer shows the actual data obtained from the experiment, whereas the bottom layer displays the modelling performed by the FNO.}
    \label{fig: rbb_1}
\end{figure}

The FNO serving as the digital twin takes in the first 10 time instances ($T_{in}$=10) along with the calibrated grid discretisations to be able to output the next time instance (step=1). The auto-regressive loop of feeding the output back along with a portion of the input and the grid discretisations is run until the $60^{th}$ time instance. The plasma takes in the first 12 ms of the plasma within the tokamak to be able to output the next 60 ms of evolution.

\textbf{Camera viewing the central solenoid :} The evolution of the plasma within the Tokamak as seen by the main camera, along with the modelling performed by the FNO can be seen in figure \ref{fig: rbb_1}. The central solenoid serves the backbone of the MAST magnet system, generating majority of the magnetic fluxes as well as the plasma current. The main camera view as shown in figure \ref{fig: elms} provides a panoramic view of the reactor, showing the poloidal cross-sectional layout on both sides of the central solenoid. The main camera has a spatial resolution of 448x640 for the experiments that we had sampled for. We had sampled a total of 59 plasma shots from the MAST database, used 50 for training and 9 for testing. The model achieved a Normalised MSE of 0.03997. 

\begin{figure}
    \centering
    \includegraphics[width=\columnwidth]{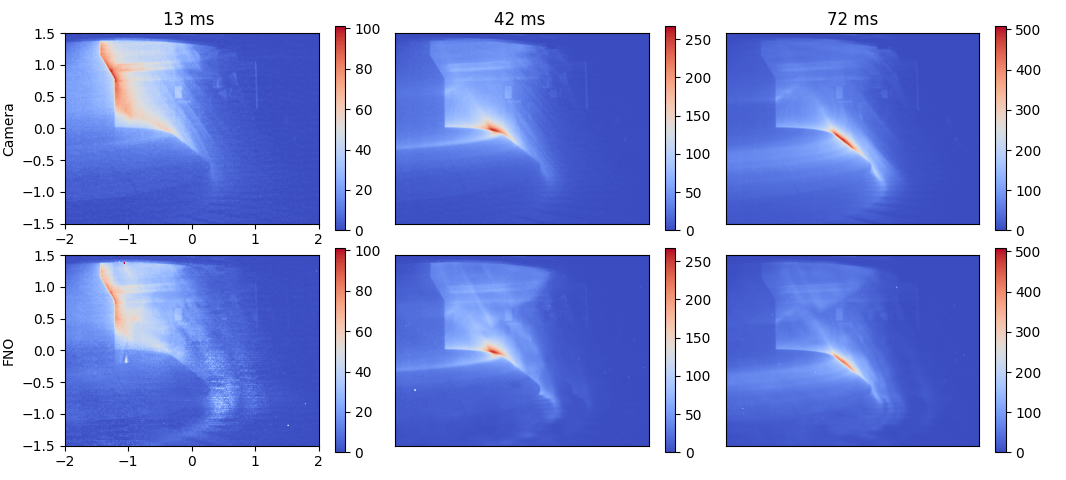}
    \caption{Temporal evolution of the plasma at the divertor within MAST as perceived by the divertor camera. The top layer shows the actual data obtained from the experiment, whereas the bottom layer displays the modelling performed by the FNO.}
    \label{fig: rba_1}
\end{figure}

\textbf{Camera viewing the divertor :} The evolution of the plasma within the Tokamak as seen by the divertor camera, along with the modelling performed by the FNO can be seen in figure \ref{fig: rba_1}. The divertor of a tokamak serves as the exhaust of the device, where the magnetic flux surfaces converge leading to the expulsion of the impurities from it. The divertor camera looks at the lower divertor of MAST with a spatial resolution of 400x512. We had sampled a total of 86 plasma shots from the MAST database, used 75 for training and 11 for testing. The model achieved a Normalised MSE of 0.01384. 

The FNO deployed for modelling the camera data accurately preserves the global features of the plasma evolution. It is able to clearly characterise the locations and intensities of the heat fluxes onto the solenoid and the divertors, allowing us to anticipate the heat flux in advance and initiate the appropriate actuator mechanism to minimise its impact onto the various plasma facing components. The FNO is also capable of modelling the filaments that arise during the plasma pulse, allowing us to predict the instances of Edge Localised Modes (ELM) that play a crucial role in the stability of the plasma. The accurate prediction of filaments also help us forecast the instance of the L-H transition, allowing us to potentially initiate the fusion reaction more efficiently. While capturing the global characteristics, the FNO performs poorly in capturing the localised phenomena with matching pixel-pixel accuracy.


\section{Conclusion and Future Work}

Through this work we demonstrate the utility of using Fourier Neural Operator as both a surrogate model and as a digital twin for modelling plasma evolution. The FNO deployed as a surrogate model is capable of solving magneto-hydrodynamics governing the transport of plasma within a fusion device, accurately and 6 orders of magnitude faster than the numerical solver. The FNO demonstrates utility in modelling each associated field with varying domain ranges. Being utilised as a digital twin, the FNO is capable of forecasting the evolution of the plasma as seen through the cameras onboard a fusion device. The FNO camera models are capable of identifying the regions of high heat deposits, filament structures, while predicting effectively the plasma evolution trajectory. The FNO for the camera can predict the plasma evolution in real-time and offers us potential in being deployed within the planning stage for various control scenarios. 

In continuing with this line of work, we will be exploring the utility of the FNO in more complex MHD cases while allowing for modelling multiple vector fields simultaneously. In this work we had chosen a MHD case with a simplified geometry, however we will be looking to expand to unstructured grids as demonstrated in \cite{Geo_FNO}. We shall also be expanding our work on the camera data with a larger dataset covering broader experimental scenarios with varying time lengths. Once proofed on a larger dataset we shall be integrating the prediction module with a plasma confinement mode classifier, allowing us to forecast the time instance of an L-H transition \cite{LH_Transition} within a tokamak. 

\newpage
\bibliographystyle{unsrt}
\bibliography{references}

\newpage

\appendix

\newpage
\section{Error Plots}
\label{appendix_errorplots}

\begin{figure}[ht]
    \centering
    \subfloat[Electric Potential]{{\includegraphics[width=0.65\textwidth]{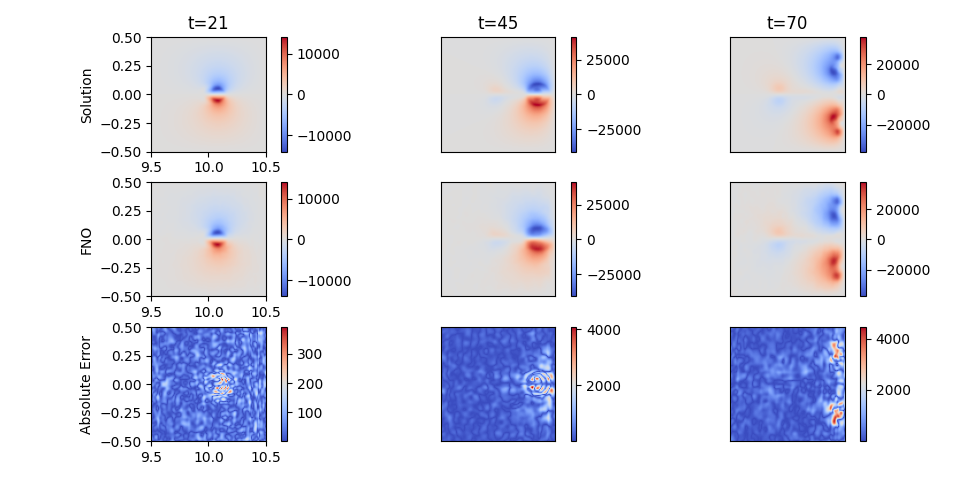}}
    \label{fig: fno_phi_2}}
    \qquad
    \subfloat[Density]{{\includegraphics[width=0.65\textwidth]{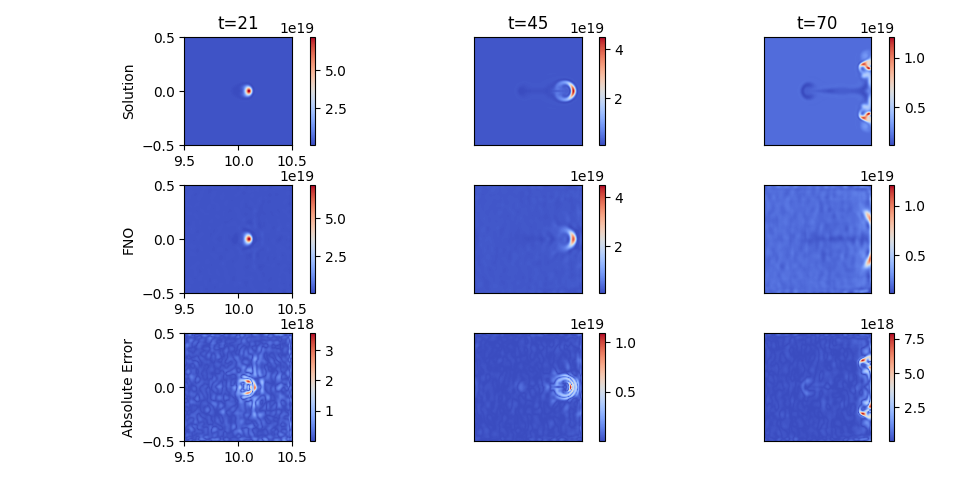}}
    \label{fig: fno_rho_2}}%
    \qquad
    \subfloat[Vorticity]{{\includegraphics[width=0.65\textwidth]{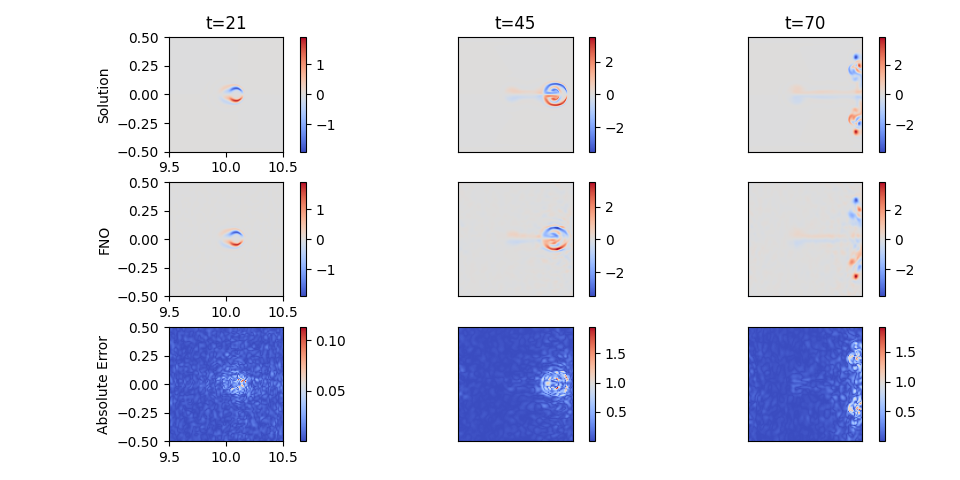}}
    \label{fig: fno_w_2}}%
    \caption{Temporal evolution of the field variables describing the plasma evolution as obtained using the code suite JOREK (top of each image) and that of the individually trained FNOs (middle of each figure) and the absolute error across both (bottom of each figures). The spatial domain is given in toroidal geometry characterised by R in the x-axis and Z in the y-axis. Figure (a) represents the temporal evolution of electric potential, figure (b) shows density and figure (c) that of the vorticity. }
    \label{fig: fno_mhd_errors}%
\end{figure}

\newpage
\section{Performance plots across various models}
\label{appendix_cross_model_performance}

\begin{figure}[ht]
    \centering
    \includegraphics[width=\textwidth]{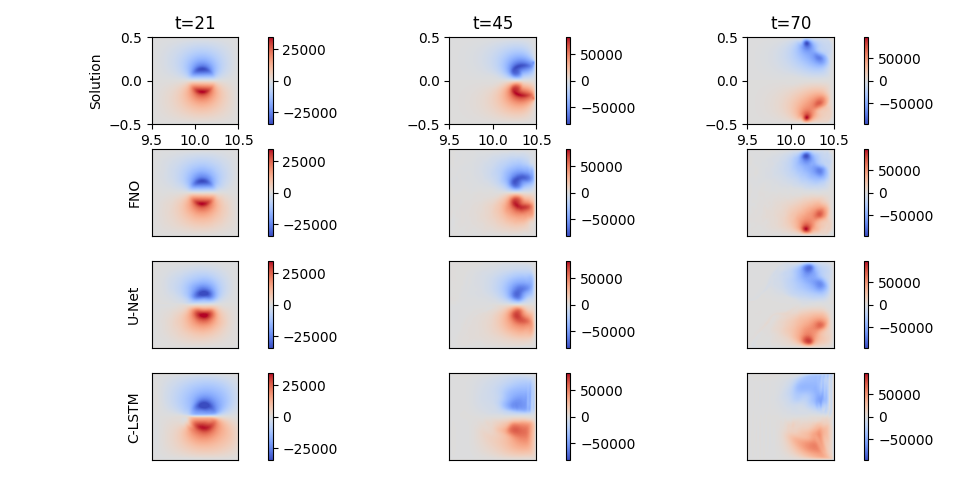}
    \caption{Evolution of the \textbf{Electric Potential} modelled by various surrogate models. Each surrogate model takes in the first 20 time instances to output in total the next 50 time instances. The Initial distribution denotes the 21$^{st}$ time instance, Middle the 45$^{th}$ and the Final 70$^{th}$ time instance. The top layer denotes the actual solution, the second layer that modelled by the FNO, the third by the U-Net and the bottom layer the output from the Conv-LSTM. }
    \label{comparison_phi}
\end{figure}

\begin{figure}
    \centering
    \includegraphics[width=\textwidth]{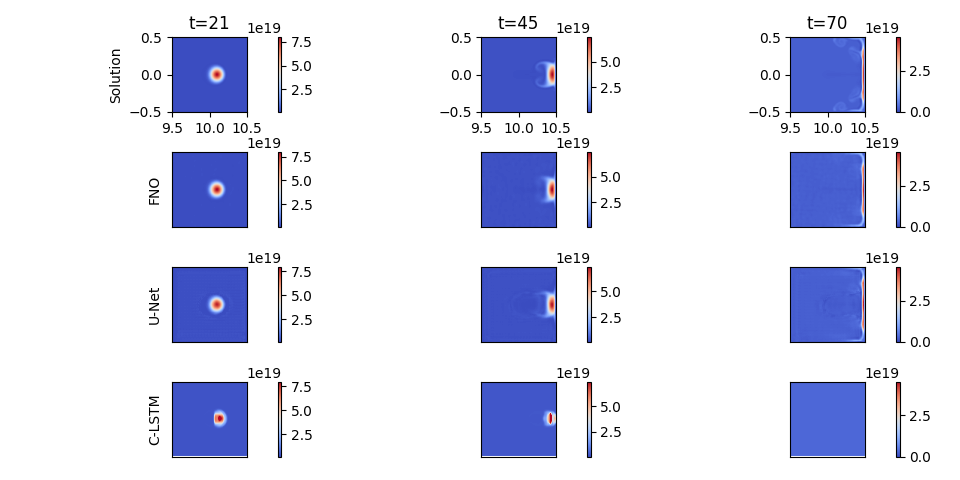}
    \caption{Evolution of the \textbf{Density} modelled by various surrogate models. Each surrogate model takes in the first 20 time instances to output in total the next 50 time instances. The Initial distribution denotes the 21$^{st}$ time instance, Middle the 45$^{th}$ and the Final 70$^{th}$ time instance. The top layer denotes the actual solution, the second layer that modelled by the FNO, the third by the U-Net and the bottom layer the output from the Conv-LSTM. }
    \label{comparison_rho}
\end{figure}

\begin{figure}
    \centering
    \includegraphics[width=\textwidth]{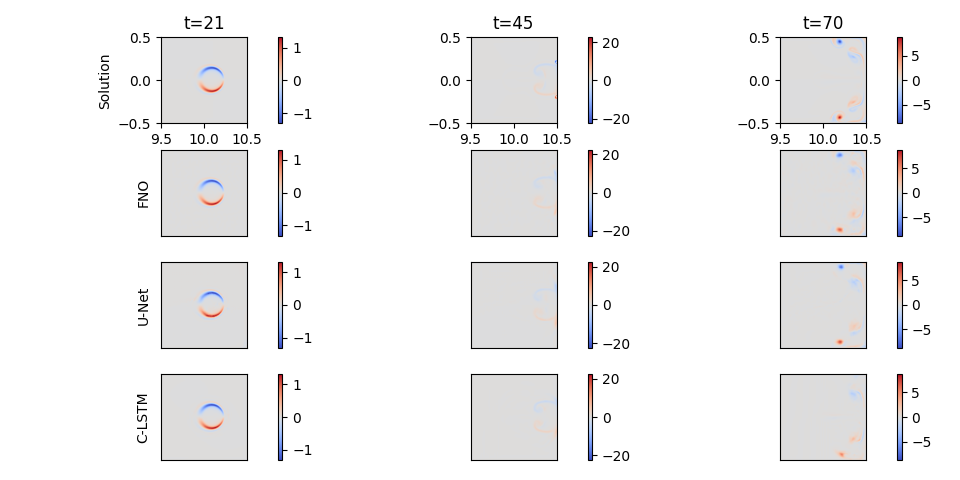}
    \caption{Evolution of the \textbf{ vorticity} modelled by various surrogate models. Each surrogate model takes in the first 20 time instances to output in total the next 50 time instances. The Initial distribution denotes the 21$^{st}$ time instance, Middle the 45$^{th}$ and the Final 70$^{th}$ time instance. The top layer denotes the actual solution, the second layer that modelled by the FNO, the third by the U-Net and the bottom layer the output from the Conv-LSTM. }
    \label{comparison_w}
\end{figure}


\begin{table}[ht]
  \caption{Comparative Performance across various Models}
  \label{table: performance_all}
  \centering
  \begin{tabular}{lllll}
    \toprule
    Model     & Field     &  Training Time (mins) & \# Parameters & NMSE  \\
    \midrule
    \rule{0pt}{2ex}    
    FNO &   & 13 & 2106981 & \textbf{4.245e-05}     \\
    ConvLSTM & Potential & 481 & 4874648 & 1.12e-03      \\
    U-Net    &  & 26 & 7931909 & 4.48e-04  \\
    \bottomrule
    \rule{0pt}{4ex}    
    FNO &  & 13 & 2106981 & \textbf{5.03e-05}     \\
    ConvLSTM & Density & 481 & 4874648 & 7.93e-03      \\
    U-Net    & & 26 & 7931909 & 1.34e-04  \\
    \bottomrule
    \rule{0pt}{4ex}    
    FNO &  & 13 & 2106981 & \textbf{4.65e-05 }    \\
    ConvLSTM & Vorticity & 479 & 4874648 & 7.98e-05      \\
    U-Net    &  & 26 & 7931909 & 5.45e-05  \\
    \bottomrule
  \end{tabular}
\end{table}

\newpage
\section{FNO 2D Architeture}
The architecture of the FNO 2D deployed in an auto-regressive manner for modelling the Reduced MHD equations. The model initially lifts the input data into a higher dimensional space using a linear layer. This is followed by 4 Fourier layers that learn the necessary modal mappings within the spatial domain, before being projected onto the lower dimensional space by a series of linear layers. \texttt{Fourier 2d} represents the 2D Fourier operator, \texttt{Conv2d} represents the 2D convolutional layer, \texttt{Add} operation adds the outputs from the Fourier layer and the convolutional layer together, \texttt{GELU} denotes the activation function: Gaussian Error Linear Unit. 
\label{appendix_fno2d}
\begin{table}[ht]
  \caption{Architecture of the FNO 2D deployed for modelling Reduced MHD}
  \label{table: fno2d}
  \centering
  \begin{tabular}{lll}
    \toprule
    Part     & Layer     &  Output Shape \\
    \midrule
    Input & - & (5, 100, 100, 22) \\    
    Lifting & \texttt{Linear} & (5, 100, 100, 32) \\
    Fourier 1 & \texttt{Fourier2d/Conv2d/Add/GELU} & (5, 32, 100, 100)\\
    Fourier 2 & \texttt{Fourier2d/Conv2d/Add/GELU} & (5, 32, 100, 100)\\
    Fourier 3 & \texttt{Fourier2d/Conv2d/Add/GELU} & (5, 32, 100, 100)\\
    Fourier 4 & \texttt{Fourier2d/Conv2d/Add/GELU} & (5, 32, 100, 100)\\
    Projection 1 & \texttt{Linear} & (5, 100, 100, 128) \\
    Projection 2 & \texttt{Linear} & (5, 100, 100, 5) \\
    \bottomrule
  \end{tabular}
\end{table}

\newpage
\section{ConvLSTM Architecture}
\label{appendix_convlstm}
The architecture of the 2D Convolutional-LSTM deployed for modelling the Reduced MHD equations. The model consists of 5 Convolutional-LSTM Cells with features scaled and lifted to 256, which is progressively scaled down to the desired output in the following manner (\texttt{[256, 128, 64, 32, 5]}). Each Conv-LSTM Cell consists of an input gate, forget gate and an output gate, where the weight matrices associated with each gate are given by 2D convolutional layers rather than being linear layers. For an understanding of the equations governing the various gates in an LSTM we refer the author to \cite{lstm}.

\begin{table}[ht]
  \caption{Architecture of the ConvLSTM deployed for modelling Reduced MHD}
  \label{table: convlstm}
  \centering
  \begin{tabular}{lll}
    \toprule
    Part     & Layer     &  Output Shape \\
    \midrule
    Input & - & (5, 20, 100, 100) \\    
    Conv-LSTM Cell 1 & \texttt{Conv2D/Sigmoid/Hadamard Product/Tanh} & (5, 256, 100, 100) \\
    Conv-LSTM Cell 2 & \texttt{Conv2D/Sigmoid/Hadamard Product/Tanh} & (5, 128, 100, 100) \\
    Conv-LSTM Cell 3 & \texttt{Conv2D/Sigmoid/Hadamard Product/Tanh} & (5, 64, 100, 100) \\
    Conv-LSTM Cell 4 & \texttt{Conv2D/Sigmoid/Hadamard Product/Tanh} & (5, 32, 100, 100) \\
    Conv-LSTM Cell 5 & \texttt{Conv2D/Sigmoid/Hadamard Product/Tanh} & (5, 5, 100, 100) \\
    \bottomrule
  \end{tabular}
\end{table}

\newpage
\section{U-Net Architecture}
\label{appendix_unet}
The architecture of the 2D U-Net deployed in an auto-regressive manner for modelling the Reduced MHD equations. The model consists of 4 encoding and 4 decoding layers. Each encoding layer is given by a 2D convolution (\texttt{Conv2d}), 2D Batch Normalisation (\texttt{BatchNorm2d}), non-linear activation (\texttt{Tanh}), followed by max-pooling layer (\texttt{MaxPool2d}). Successive to the encoding layers, we establish a bottle neck layer which is identical to an encoding layer but without the pooling element. In each Decoder, the information is up-scaled via a 2D transposed convolution (\texttt{ConvTranspose2D}), the output of which is concatenated with that of the output of the corresponding encoder and then passed through a series of 2D convolutional layers. The final output is rescaled to the desired dimensions using 2D convolutions.  
\begin{table}[ht]
  \caption{Architecture of the 2D U-Net deployed for modelling Reduced MHD}
  \label{table: unet}
  \centering
  \begin{tabular}{lll}
    \toprule
    Part     & Layer     &  Output Shape \\
    \midrule
    Input & - & (5, 20, 100, 100) \\    
    Encoder 1 & \texttt{Conv2d/BatchNorm2d/Tanh} & (5, 32, 100, 100) \\
    Pool 1 & \texttt{MaxPool2d} & (5, 32, 50, 50)\\
    Encoder 2 & \texttt{Conv2d/BatchNorm2d/Tanh} & (5, 64, 50, 50) \\
    Pool 2 & \texttt{MaxPool2d} & (5, 64, 25, 25)\\
    Encoder 3 & \texttt{Conv2d/BatchNorm2d/Tanh} & (5, 128, 25, 25) \\
    Pool 3 & \texttt{MaxPool2d} & (5, 128, 12, 12)\\
    Encoder 4 & \texttt{Conv2d/BatchNorm2d/Tanh} & (5, 256, 12, 12) \\
    Pool 4 & \texttt{MaxPool2d} & (5, 256, 6, 6)\\
    Bottleneck & \texttt{Conv2d/BatchNorm2d/Tanh} & (5, 512, 6, 6) \\
    Decoder 4 & \texttt{ConvTranspose2D/Encoder 4} & (5, 256, 12, 12) \\
    Decoder 3 & \texttt{ConvTranspose2D/Encoder 3} & (5, 128, 25, 25) \\
    Decoder 2 & \texttt{ConvTranspose2D/Encoder 2} & (5, 64, 50, 50) \\
    Decoder 1 & \texttt{ConvTranspose2D/Encoder 1} & (5, 32, 100, 100) \\
    Rescale  & \texttt{Conv2D} & (5, 20, 100, 100) \\
    \bottomrule
  \end{tabular}
\end{table}

\end{document}